\documentclass[lettersize,journal]{IEEEtran}
\usepackage{amsmath,amsfonts}
\usepackage{algorithmic}
\usepackage{array}
\usepackage[caption=false,font=normalsize,labelfont=sf,textfont=sf]{subfig}
\usepackage{textcomp}
\usepackage{stfloats}
\usepackage{url}
\usepackage{verbatim}
\usepackage{graphicx}

\usepackage{float} 
\usepackage{amssymb,mathrsfs,amsthm,amsbsy,graphicx,color,bm,balance,mathtools,xfrac,nccmath,siunitx,cite,soul}

\usepackage{xcolor}
\usepackage{array} 
\usepackage{makecell}
\usepackage{tabularx}
\usepackage{bm}
\usepackage{siunitx, soul}

\newtheorem{rem}{Remark}

\usepackage{pgfplots}
\pgfplotsset{compat=newest}
\usetikzlibrary{plotmarks}
\usetikzlibrary{arrows.meta}
\usepgfplotslibrary {patchplots}
\usepackage{grffile}

\newcommand{\thr}{\rm{th}}

\newcommand{\qh}{\mathbf{h}}
\newcommand{\qa}{\mathbf{a}}
\newcommand{\qf}{\mathbf{f}}
\newcommand{\qs}{\mathbf{s}}
\newcommand{\qg}{\mathbf{g}}

\newcommand{\qv}{\mathbf{v}}

\newcommand{\qw}{\mathbf{w}}
\newcommand{\qx}{\mathbf{x}}

\newcommand{\qE}{\mathbf{E}}

\newcommand{\qS}{\mathbf{S}}

\newcommand{\qW}{\mathbf{W}}

\newcommand{\qI}{\mathbf{I}}

\newcommand{\qV}{\mathbf{V}}

\newcommand{\tr}{\text{Tr}}

\def\BibTeX{{\rm B\kern-.05em{\sc i\kern-.025em b}\kern-.08em
    T\kern-.1667em\lower.7ex\hbox{E}\kern-.125emX}}
\usepackage{balance}
\begin{document}
\bstctlcite{IEEEexample:BSTcontrol}

\title{{\huge Low-Complexity   Multi-Target Detection in  ELAA ISAC}\vspace{-0mm}}
\author{Diluka  Galappaththige, \IEEEmembership{Member, IEEE},  Shayan Zargari, Chintha Tellambura, \IEEEmembership{Fellow, IEEE,} and Geoffrey Ye Li, \IEEEmembership{Fellow, IEEE,}
\thanks{D.  Galappaththige, S. Zargari, and C. Tellambura are with the Department of Electrical and Computer Engineering, University of Alberta, Edmonton, AB, T6G 1H9, Canada (e-mail: \{diluka. lg, zargari, ct4\}@ualberta.ca).\\
 \indent G. Y. Li is with the ITP Lab, the Department of Electrical and Electronic Engineering, Imperial College London, SW7 2BX London, U.K.(e-mail: geoffrey.li@imperial.ac.uk).} \vspace{-3mm}}

\markboth{Journal of \LaTeX\ Class Files,~Vol.~18, No.~9, September~2020}%
{How to Use the IEEEtran \LaTeX \ Templates}

\maketitle

\begin{abstract}
Multi-target detection and communication with extremely large-scale antenna arrays (ELAAs) operating at high frequencies necessitate generating multiple beams. However, conventional algorithms are slow and computationally intensive.  For instance, they can simulate a  \num{200}-antenna system over two weeks, and the time complexity grows exponentially with the number of antennas. Thus, this letter explores an ultra-low-complex solution for a multi-user, multi-target integrated sensing and communication (ISAC) system equipped with an ELAA base station (BS). It maximizes the communication sum rate while meeting sensing beampattern gain targets and transmit power constraints. As this problem is non-convex, a Riemannian stochastic gradient descent-based augmented Lagrangian manifold optimization (SGALM) algorithm is developed, which searches on a manifold to ensure constraint compliance. The algorithm achieves ultra-low complexity and superior runtime performance compared to conventional algorithms.  For example, it is \num{56} times faster than the standard benchmark for \num{257} BS antennas.
\end{abstract}


\section{Introduction}
\IEEEPARstart{M}{ulti}-target detection in integrated sensing and communication (ISAC)  systems is a critical application. However, ISAC base stations (BSs) will use large-scale antenna arrays (ELAA) at high frequencies  \cite{Wang2024, Diluka2024NF}, which exhibit near-field (NF) propagation \cite{Azar2024}. Of course, far-field (FF) ISAC research is also emerging \cite{Sun2024, Sun2024MultiFunctional, zargari2024CFISAC}. The high dimensionality of ELAA systems,  e.g., a vast number of antenna elements, exponentially increases the computational complexity  \cite{Wang2024, Diluka2024NF}. Traditional successive convex approximation (SCA) and semi-definite relaxation (SDR) methods are inefficient due to overhead and extended running times. For example, in \cite{Diluka2024NF}, a 200-antenna BS ISAC system required two weeks to generate a single simulation curve, with time complexity growing exponentially as the number of antennas increases.

This challenge calls for super-efficient algorithms. The problem's search space has size $MKN$, where $M$, $K$, and $N$ are the numbers of BS antennas, users, and targets, respectively. Consequently, the search space grows rapidly as network dimensions increase, leading to exponential complexity. Thus, a fundamentally different algorithm is required.

{The key novelty of our work lies not only in the algorithm but in addressing the high computational complexity of beamforming for multi-user, multi-target NF ISAC systems with ELAAs. To our knowledge, multi-target detection in NF ISAC systems remains largely unexplored. The closest related work, our previous paper {\cite{Diluka2024NF}}, employs SCA and SDR for FF ISAC systems.  This study is the first to introduce a low-complexity beamforming for multi-target detection and communication in NF ISAC systems. In contrast, many NF ISAC studies focus on single-target scenarios. }

{Building on {\cite{zargari2024riemannian}}, our algorithm extends beyond standard optimization (including manifold optimization (MO) and fractional programming (FP)) by integrating a penalty-based augmented Lagrangian method (ALM) directly into the MO framework. While {\cite{zargari2024riemannian}} developed a low-complexity iterative augmented Lagrangian MO (IALMO) algorithm for FF ISAC systems, it focuses on a {\qty{3}{\GHz}} frequency with fewer than $28$ BS antennas. At higher frequencies, like {\qty{54}{\GHz}}, smaller wavelengths enable antenna arrays with up to $500$ elements. Although adaptable, IALMO's performance has not yet been evaluated for ELAA-ISAC systems over the {\qty{10}{\GHz}}-{\qty{10}{\THz}} range. Moreover, {\cite{zargari2024riemannian}} reuses the same communication beams for target detection. In NF, this dual usage is impractical due to beam-focusing effects {\cite{Azar2024}}, where energy is concentrated in both angle and distance, reducing coverage and limiting multi-purpose applications. To address this, our approach employs separate beams for communication and sensing (C\&S)  with ELAA (see Remark~{\ref{rem_beamfocusing}}).}

{Unlike {\cite{zargari2024riemannian}}, which uses Riemannian conjugate (RC) methods, we adopt Riemannian stochastic gradient descent (SGD) to solve the proposed beamforming problem. Riemannian SGD improves large-scale optimization by approximating gradients using subsets of data at each iteration. This reduces computational overhead compared to full gradient descent, allowing faster updates and enhanced scalability {\cite{Bonnabel6487381}}.}

This letter presents a Riemannian SGD-based augmented Lagrangian MO (SGALM) algorithm to overcome the aforementioned computational complexity.  It is designed to maximize communication sum rates while meeting sensing thresholds and minimum rate requirements.  Since the optimization problem is non-convex and NP-hard, we augment the objective function with a penalty for beampattern gain violations and minimize over a complex sphere manifold, following the ALM \cite{liu2020simple}. SGALM iteratively adjusts optimization variables and Lagrange parameters to satisfy constraints. Compared to standard techniques, SGALM archives dramatic complexity savings. For example, with $257$ and $321$ BS antennas, SGALM is $56$ and $61$ times faster than the benchmark. This efficiency gain scales up with the antenna array size, making SGALM ideal for ELAA-ISAC systems and dynamic networks.

While this study focuses on NF propagation, SGALM is versatile and can be applied to various operating scenarios, including FF environments and different antenna configurations. This flexibility makes it suitable for diverse applications, ranging from urban environments with dense multi-path interference to rural areas with line-of-sight conditions.

\section{Preliminaries}\label{system_modelA}

\begin{figure}[!t]\vspace{-0mm}
    \centering 
    \def\svgwidth{180pt} 
    \fontsize{7}{7}\selectfont 
    \graphicspath{{Figures/}}
    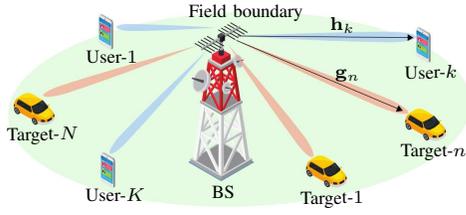 \vspace{-3mm} 
    \caption{A ELAA-ISAC system.}  \label{fig_SystemModel}\vspace{-3mm}
\end{figure}

\subsubsection{System and Channel Models}
Fig.~\ref{fig_SystemModel} shows the narrowband ELAA-ISAC system consisting of $M=2\tilde{M}+1$-uniform linear array (ULA) antenna BS with antenna spacing $d = \lambda/2$, $K$ single-antenna communication users, and $N$ sensing targets, where $\lambda$ is the wavelength  \cite{Diluka2024NF}. The NF spherical wave channel model is used. The NF channel vector between the BS and the $b$-th node ($b=k$ for the $k$-th user and $b=n$ for the $n$-th target), $\qf_b \in \mathbb{C}^{M\times1}$, is given as $\qf_b = \beta_b \qa_f (r_b, \theta_b),$ where  $\qf\in\{\qh,\qg\}$, $\qa_f (r_b, \theta_b) \in \mathbb{C}^{M\times1}$ is the NF array response vector with $[\qa_f (r_b, \theta_b)]_{m} = e^{-j\frac{2\pi}{\lambda} (r_{mb}(r_b,\theta_b) - r_b)}$ as the $m$-th element, and $r_{mb}(r_b,\theta_b)$ is given in \cite[Eqn. (1)]{Diluka2024NF}. Here, $r_b$ and $\theta_b$ are the $b$-th node distance and angle from the center of the ULA, respectively. Moreover, $\beta_{b} = \tilde{\beta}_{b} e^{-j\frac{2\pi}{\lambda} r_b}$ is the complex channel gain, and $\tilde{\beta}_{b} = \sqrt{{\lambda}/{4\pi}}r_b^{-1}$ is the free-space path-loss between the $0$-th antenna element and the $b$-th node. Further details of the channel and system models can be found in \cite{Diluka2024NF}. 

Channel estimation and data transmission/sensing occur in two separate time slots. The initial slot in each coherent interval is allocated to estimating channel state information (CSI), feasible with emerging methods \cite{Zhang2022Fast}. Thus,  CSI is available. This assumption is standard and widely used. {Nevertheless, the proposed beamforming algorithm is robust to imperfect CSI. The details of how it can adapt to imperfect CSI can be found in  {\cite{zargari2024riemannian}}.}

\subsubsection{Signal Model}
The BS transmits signal $\qx\in \mathbb{C}^{M\times 1}$, which is jointly designed for communicating with users while sensing the targets, i.e, $\qx = \sum_{k \in \mathcal{K}} \qw_{k} q_{k} + \sum_{n \in \mathcal{N}}\qs_n$. Here, $q_{k} \in \mathbb{C}$ denotes the intended data symbol for the $k$-th user with unit power, i.e., $ \mathbb{E}\{\vert q_{k} \vert^2 \}=1$, $\qw_{k} \in \mathbb{C}^{M\times 1}$ is the BS data beamforming vector for the $k$-th user, and $\mathbf{s}_n \in \mathbb{C}^{M\times 1}$ is the sensing signal intended for detecting the $n$-th target. It is assumed that $q_{k}$ and $\qs_n$ are independent of each other for $k\in\mathcal{K}$ and $n\in \mathcal{N}$, and the beamforming at the BS is achieved through designing $\qw_{k}$ and $\qs_n$ \cite{Zhenyao2023, Diluka2023}. 
The received signal at the $k$-th user is given by
\begin{eqnarray}\label{eqn_rx_user_k}
    y_{k} = \qh_{k}^{\rm{H}} \qw_{k} q_{k} + \sum\nolimits_{i \in \mathcal{K}_k} \qh_{k}^{\rm{H}} \qw_{i} q_{i} + \sum\nolimits_{n\in \mathcal{N}} \qh_k^{\rm{H}} \qs_n + \varsigma_{k},\quad
\end{eqnarray}
where $\varsigma_k\sim \mathcal{CN}(0,\sigma^2)$ denotes the additive white Gaussian noise (AWGN) at the $k$-th user.

\begin{rem}\label{rem_beamfocusing}
    Multiple $N$-independent sensing beams are employed to detect $N$ separate targets. Unlike FF ISAC systems, NF transmit beams are concentrated in both direction (angle) and distance, resulting in narrow beams (the beam-focusing effect). It is thus necessary to concurrently generate multiple beams to achieve improved sensing performance \cite{Cong2024}.  
\end{rem}

\textbf{Mathematical notations:} The beamforming vectors are organized into a single matrix $\qV = [\qw_1, \ldots, \qw_K, \qs_1, \ldots, \qs_N] \in \mathbb{C}^{M\times (K+N)}$. An index matrix, $\qE = \qI_{K+N} \in \mathbb{R}^{(K+N)\times(K+N)}$, is defined. The combination of $\qV$ and $\qE$ allows the representation of each column individually from $\qV$.

\section{Communication and Sensing Performance}
Communication rates at the users and the transmit beam patterns for the targets at the BS are determined. 

\subsubsection{Communication Rates}
Users decode their data using the BS  signal. From \eqref{eqn_rx_user_k}, the received SINR at the $k$-th user is given by 
\begin{equation}\label{eqn_gamma_bar}
    \gamma_k=\frac{\vert \qh_k^{\rm{H}} \qV \qE_{k}\vert^2}{\sum_{i\in \mathcal{L}_k} \vert \qh_k^{\rm{H}} \qV \qE_{i} \vert^2 + \sigma^2},~\forall k,
\end{equation}
where $\qE_k$ is the $k$-th column of $\qE$. The  $k$-th user rate can be approximated  as $\mathcal{R}_{k} = \log_2(1+ \gamma_{k})$. 

\subsubsection{Sensing Beampattern}
The transmit power can vary as a function of sensing angle $\theta\in[-\pi/2, \pi/2]$, enhancing detection, range/Doppler/angle estimation, tracking, recognition, and sensing accuracy \cite{Stoica2007}.   The sensing beampattern  can be defined at the $n$-th target direction as 
\begin{eqnarray}\label{qen_nT_BG_bar}
    p(\theta_{n}) &=& \mathbb{E}\left\{ |\qg_n^{\rm{H}} \qx |^2 \right\} = \sum\nolimits_{i\in\mathcal{K}}  \qg_n^{\rm{H}} \qV \qE_{i} \qE_{i}^{\rm{H}} \qV^{\rm{H}} \qg_n \nonumber \\
    && + \sum\nolimits_{j\in\mathcal{N}} \qg_n^{\rm{H}} \qV \qE_{K+j} \qE_{K+j}^{\rm{H}} \qV^{\rm{H}} \qg_n,~\forall n.
\end{eqnarray}
This measure is tailored to meet specific target sensing requirements. For instance, when the directions of potential targets are unknown, a uniformly distributed $p(\theta) $ is the most effective choice. Conversely, in applications such as target tracking, where the potential directions of targets are approximately known, the measure should be designed to enhance the gain in those specific directions, thereby improving the accuracy and efficiency of target identification and tracking \cite{Stoica2007}.

\section{Problem Formulation}\label{Sec_prob_form}
The primary goal is to maximize the users' communication sum rate while adhering to constraints on $p(\theta_n)$, minimum rate demands, and maximum BS transmit power. The beamforming vectors are optimized as follows: 
\begin{subequations}\label{eqn_P}
\begin{align}
    \mathbf{P}:~& \max_{\qV } \quad  \sum\nolimits_{k\in \mathcal{K}} \log_2(1 + \gamma_k), \label{eqn_P_obj}  \\
     \text{s.t.} \quad &  p(\theta_{n})  \geq \Omega_{n}^{\thr}, ~\forall n,\label{eqn_P_beamgain}  \\
     & \gamma_k \geq \Gamma_k^{\thr}, ~\forall k,\label{eqn_P_comm_rate} \\
    & \tr(\qV \qV^{\rm{H}}) \leq p_{\rm{max}}, \label{eqn_P_tx_pow} 
\end{align}
\end{subequations}
where \eqref{eqn_P_beamgain} ensures the sensing beampattern gain required for each target, in which $\Omega_{n}^{\thr}$ is the intended sensing beampattern gain for the $n$-th target, {{\eqref{eqn_P_comm_rate}} is the SINR constraint at each user with  $\Gamma_k^{\thr} = 2^{\mathcal{R}_k^{\thr}}-1$ where $\mathcal{R}_k^{\thr}$ is the $k$-th user rate requirement,} and \eqref{eqn_P_tx_pow} sets the BS transmit power constraint, with a maximum allowable transmit power of $p_{\rm{max}}$.

Note that the problem formulation in $\mathbf{P}$ and \cite{Diluka2024NF} differ in various aspects. In particular, they have distinct objectives, i.e., \cite{Diluka2024NF} minimizes the BS transmit power under C\&S rate constraints, whereas $\mathbf{P}$ maximizes the communication sum rate while considering sensing beampattern gain and BS transmit power. Conversely, \cite{Diluka2024NF} and $\mathbf{P}$ use two different sensing performance measures, i.e., sensing rate and beampattern gain. The sensing rate utilizes transmit and receiver beampatterns, reflecting how much environmental information can be collected from a target's reflected signal \cite{Zhenyao2023, Diluka2023}. However, it requires the BS to function in full-duplex mode, severely compromising performance due to self-interference. Here, beampattern gain is used as a viable performance metric \cite{Stoica2007}. Nonetheless, our method to address $\mathbf{P}$ can easily accommodate the sensing rate.

\section{Proposed Solution}\label{Sec_proposed_solution}
Non-convex optimization problem $\mathbf{P}$ is solved using MO and FP to determine the optimal BS transmit beamforming vectors. 
To tackle the challenges imposed by the sum-log terms in $\mathbf{P}$,  the FP approach is used to substitute auxiliary variables, $\boldsymbol{\mu} = [\mu_1, \ldots, \mu_K]$, for each SINR term in \eqref{eqn_P_obj} while ensuring $\mu_k \leq {\gamma}_k$. Thereby, $\mathbf{P}$ is equivalently reformulated as \cite{Shen2018}
\begin{subequations}\label{eqn_P2}
\begin{align}
    \mathbf{P}_1:~& \max_{\qV,\boldsymbol{\mu}} ~f(\qV, \boldsymbol{\mu}) = \frac{1}{\ln(2)} \sum\nolimits_{k\in \mathcal{K}} \ln(1 + \mu_k)  \nonumber\\
    &\hspace{7mm} + \frac{1}{\ln(2)} \sum\nolimits_{k\in \mathcal{K}} \left( - \mu_k + \frac{(1 + \mu_k){\gamma}_k}{1 + {\gamma}_k} \right), \label{eqn_P2_obj}  \\
    \text{s.t.} \quad &  \eqref{eqn_P_beamgain}-\eqref{eqn_P_tx_pow}.
\end{align}
\end{subequations}
Note that $\mathbf{P}_1$ is a two-part optimization problem: (i) an outer optimization over $\qV$ with fixed $\boldsymbol{\mu}$ and (ii) an inner optimization over $\boldsymbol{\mu}$ with fixed $\qV$ \cite{Shen2018}. To solve $\mathbf{P}_1$,  $\qV$ and $\boldsymbol{\mu}$ are alternatively optimized until the objective function converges.

\subsubsection{Optimizing $\boldsymbol{\mu}$ for Fixed $\qV$}
For a given $\qV$, $f(\qV, \boldsymbol{\mu})$ is a concave differentiable function over $\boldsymbol{\mu}$. Consequently, the optimal $\boldsymbol{\mu}$ is computed by setting each $\frac{\partial f(\qV, \boldsymbol{\mu})}{\partial \mu_k}$ to zero. The optimal $\mu_k$  is given by $\mu_k^* ={\gamma}_k$. Note that substituting $\boldsymbol{\mu}^*$ back into $f(\qV, \boldsymbol{\mu})$ yields the exact sum-of-logarithms objective function in $\mathbf{P}$.

\subsubsection{Optimizing $\qV$ for Fixed $\boldsymbol{\mu}$}
For a given $\boldsymbol{\mu}$,  the objective \eqref{eqn_P2_obj} is simplified, eliminating the constant terms with respect to $\qV$, and reformulate $\mathbf{P}_1$ as
\begin{subequations}\label{eqn_P3}
\begin{align}
    \!\!\!\mathbf{P}_2: \max_{\qV} \,  \sum_{k\in \mathcal{K}}  \frac{\hat{\mu}_k \vert  \qh_k^{\rm{H}} \qV \qE_{k}\vert^2}{\sum_{i\in \mathcal{L}} \vert \qh_k^{\rm{H}} \qV \qE_{i} \vert^2 + \sigma^2} , \,\,\,  \text{s.t.} \,\, \eqref{eqn_P_beamgain}-\eqref{eqn_P_tx_pow}, 
\end{align}
\end{subequations}
where $\hat{\mu}_k = 1+\mu_k$. Note that $\mathbf{P}_2$ maintains equivalence with original problem $\mathbf{P}$, resulting in no performance loss \cite{zargari2024riemannian}.

An efficient approach is proposed to address $\mathbf{P}_2$ that leverages MO to obtain optimal transmit beamforming. MO restricts the search space to a manifold, which locally resembles Euclidean spaces. MO algorithms thus efficiently navigate this space by leveraging its geometric properties to find optimal solutions \cite{liu2020simple}. For $\mathbf{P}_2$, the relevant manifold is a subspace with dimensions $(M+1)(K+N)$, resulting in significant reductions in computational complexity. Moreover, searching on a manifold eliminates the need for relaxations or approximations, which are major drawbacks of conventional optimization techniques. This allows MO algorithms to achieve more accurate solutions with greater efficiency.

First, the power constraint is normalized to ensure that $\text{Tr}(\qV \qV^{\rm{H}}) \leq 1$. Then, a modified matrix $\tilde{\qV}$ is introduced,  composed by columns $\{\tilde{\qv}_1, \ldots, \tilde{\qv}_L\}$, which satisfies the condition $\text{Tr}(\tilde{\qV} \tilde{\qV}^{\rm{H}}) = \text{Tr}(\qV \qV^{\rm{H}}) + ||\mathbf{z}||_2^2=1$, where $\tilde{\qv}_k = [\qw_k^{\rm{T}}, z_k]^{\rm{T}}$ for $k\in \mathcal{K}$ and $\tilde{\qv}_n = [\qs_n^{\rm{T}}, z_n]^{\rm{T}}$ for $n\in \{K+1, \ldots, K+N\}$. Here, $\mathbf{z} = [z_1, \ldots, z_{K+N}]$ is an auxiliary vector introduced to simplify power normalization while preserving the constraint. This leads to a complex sphere manifold as $\mathcal{M} = \{ \tilde{\qV} \in \mathbb{C}^{(M+1) \times (K+N)} \:|\: \text{Tr}(\tilde{\qV} \tilde{\qV}^{\rm{H}}) = 1 \}$. Thus, $\mathbf{P}_2$ can be transformed into a constrained optimization problem on $\mathcal{M}$ as
\begin{subequations}\label{eqn_P4}
\begin{eqnarray}
     \mathbf{P}_3:&& \!\!\! \min_{\tilde{\qV} \in \mathcal{M}} \, f(\tilde{\qV}) = -\!\sum_{k\in \mathcal{K}} \frac{\hat{\mu}_k \vert  \hat{\qh}_k^{\rm{H}} \tilde{\qV}  \qE_{k}\vert^2}{\sum_{i\in \mathcal{L}} \vert \hat{\qh}_k^{\rm{H}} \tilde{\qV}  \qE_{i} \vert^2 + \sigma^2} ,  \\
    \text{s.t.} \,\, &&  u_n(\tilde{\qV})  =\Omega_{n}^{\thr} - \sum\nolimits_{i\in\mathcal{K}}  \hat{\qg}_n^{\rm{H}} \tilde{\qV} \qE_{i} \qE_{i}^{\rm{H}} \tilde{\qV}^{\rm{H}} \hat{\qg}_n \nonumber \\
    &&\quad- \sum\nolimits_{j\in\mathcal{N}} \hat{\qg}_n^{\rm{H}} \tilde{\qV} \qE_{K+j} \qE_{K+j}^{\rm{H}} \tilde{\qV}^{\rm{H}} \hat{\qg}_n \leq 0 ,~\forall n, \qquad \label{eqn_P4_sen} \\
    && c_k(\tilde{\qV}) \!=\!\Gamma_{k}^{\thr} - \frac{\vert  \mathbf{\hat{h}}_k^{\rm{H}} \tilde{\qV}  \qE_{k}\vert^2}{\sum_{i\in \mathcal{L}_k} \vert \mathbf{\hat{h}}_k^{\rm{H}} \tilde{\qV}  \qE_{i} \vert^2 + \sigma^2}\leq 0, \forall k, \label{eqn_P4_sinr}
\end{eqnarray}
\end{subequations}
where $\hat{\qh}_k = \sqrt{p_{\rm{max}}}[\qh_k, 0]$ and $\hat{\qg}_n = \sqrt{p_{\rm{max}}}[\qg_n, 0]$ are adjusted to match the problem's dimensionality. In  $\mathcal{M}$, $f(\tilde{\qV})$ and  $u_n(\tilde{\qV})$ are continuous differentiable functions from $\mathcal{M}$ to $\mathbb{R}$. However, $\mathbf{P}_3$ involves constraint \eqref{eqn_P4_sen}, which is beyond the manifold constraint.  Fortunately,  constraint \eqref{eqn_P4_sen} can be incorporated into the objective as a penalty term \cite{liu2020simple}. Thus, the resulting Lagrangian function is given as follows \cite{liu2020simple}:
\begin{eqnarray}\label{Lag_penalty}
    \mathcal{L}_\rho(\tilde{\qV} , \boldsymbol{\lambda}, \boldsymbol{\kappa}) &&= f(\tilde{\qV}) + \frac{\rho}{2} \sum\nolimits_{n \in \mathcal{N}} \max\left\{0, \frac{\lambda_n}{\rho} + {u_n(\tilde{\qV})}\right\}^2 \nonumber \\
    && + \frac{\rho}{2} \sum\nolimits_{k \in \mathcal{K}} \max\left\{0, \frac{\kappa_k}{\rho} + c_k(\tilde{\qV})\right\}^2,
\end{eqnarray}
{where $\rho > 0$ is a penalty parameter and $\boldsymbol{\lambda} \geq 0 \in \mathbb{R}^{N}$ and $\boldsymbol{\kappa} \geq 0 \in \mathbb{R}^{K}$ are the vectors of Lagrange parameters. The ALM optimizes $\tilde{\qV}$ for a given $\boldsymbol{\lambda}$ and $\boldsymbol{\kappa}$ utilizing the MO approach and updates $\boldsymbol{\lambda}$ and $\boldsymbol{\kappa}$ with a gradient-type rule {\cite{Birgin2014book}}. }

{Interested readers are referred to {\cite{liu2020simple, zargari2024riemannian}} for the optimality and convergence properties of the proposed algorithm.} To optimize \eqref{Lag_penalty} on $\mathcal{M}$, the SGALM algorithm involves the following main steps \cite{liu2020simple, zargari2024riemannian}.

\textbf{Riemannian SGD}: The Riemannian SDG is similar to the deterministic case, but the Euclidean gradient $\nabla_{\tilde{\qV}_t} \mathcal{L}_\rho(\tilde{\qV}, \boldsymbol{\lambda}, \boldsymbol{\kappa})$ is computed on a mini-batch of data or stochastic samples at each iteration. This stochastic Euclidean gradient is projected onto the tangent space $T_{\mathbf{\tilde{V}}_t}\mathcal{M}$.
The projection results as follows: ${\rm{grad}}_{\tilde{\qV}_t} \mathcal{L}_\rho(\tilde{\qV}, \boldsymbol{\lambda}, \boldsymbol{\kappa}) = \nabla_{\tilde{\qV}_t} \mathcal{L}_\rho(\tilde{\qV}, \boldsymbol{\lambda}, \boldsymbol{\kappa}) - \Re\{\nabla_{\tilde{\qV}_t} \mathcal{L}_\rho(\tilde{\qV} , \boldsymbol{\lambda}, \boldsymbol{\kappa}) \circ \tilde{\qV}^*_t\}\circ \tilde{\qV}_t$.
This ensures the update stays on $\mathcal{M}$. The Euclidean gradient of the objective function \eqref{Lag_penalty} is given by \eqref{derivtive_eq}.

\begin{figure*}[!t]
\begin{small}
\begin{align}  \label{derivtive_eq}
    \nabla_{\tilde{\qV}_t} \mathcal{L}_\rho(\tilde{\qV} , \boldsymbol{\lambda}, \boldsymbol{\kappa}) & =  \sum\nolimits_{k\in \mathcal{K}} -\hat{\mu}_k  \Biggl( \frac{2  \hat{\qh}_k^{\rm{H}} \tilde{\qV}_t \mathbf{E}_{k} \hat{\qh}_k \mathbf{E}_{k}^{\rm{H}} }{\sum_{j\in \mathcal{L}} \vert\hat{\qh}_k^{\rm{H}} \tilde{\qV}_t \mathbf{E}_{j}\vert^2 + \sigma^2} -   \sum_{i\in \mathcal{L}} \frac{2 \vert\hat{\qh}_k^{\rm{H}} \tilde{\qV}_t \mathbf{E}_{k}\vert^2  \hat{\qh}_k^{\rm{H}} \tilde{\qV}_t \mathbf{E}_{i}\hat{\qh}_k\mathbf{E}_{i}^{\rm{H}}  }{\left(\sum_{j\in \mathcal{L}} \vert\hat{\qh}_k^{\rm{H}} \tilde{\qV}_t \mathbf{E}_{j}\vert^2 + \sigma^2\right)^2} \Biggl) \nonumber \\
    & \!\!\!\!\!\!\!\!- 2 \rho \sum\nolimits_{n \in \mathcal{N}}   \mathbf{1}_{\left\{\lambda_n/{\rho} + {u_n(\tilde{\qV})}\right\}} \left({\lambda_n}/{\rho} + {u_n(\tilde{\qV})}\right) \left(\sum\nolimits_{i\in\mathcal{K}} \hat{\qg}_n^{\rm{H}} \tilde{\qV} \qE_{i} \hat{\qg}_n \qE_{i}^{\rm{H}} + \sum\nolimits_{j\in\mathcal{N}} \hat{\qg}_n^{\rm{H}} \tilde{\qV} \qE_{K+j} \hat{\qg}_n \qE_{K+j}^{\rm{H}}  \right) \nonumber \\
    &\!\!\!\!\!\!\!\! - 2 \rho \sum\nolimits_{k \in \mathcal{K}}   \mathbf{1}_{\left\{{\kappa_k}/{\rho} + {c_k(\tilde{\qV})}\right\}} \left({\kappa_k}/{\rho} + {c_k(\tilde{\qV})}\right) \Biggl( \frac{2\mathbf{\hat{h}}_k^{\rm{H}} \tilde{\qV}_t \mathbf{E}_{k} \mathbf{\hat{h}}_k \mathbf{E}_{k}^{\rm{H}} }{\sum_{j\in \mathcal{L}_k} |\mathbf{\hat{h}}_k^{\rm{H}} \tilde{\qV}_t \mathbf{E}_{j}|^2 + \sigma^2} -   \sum_{i\in \mathcal{L}_k} \frac{2|\mathbf{\hat{h}}_k^{\rm{H}} \tilde{\qV}_t \mathbf{E}_{k}|^2  \mathbf{\hat{h}}_k^{\rm{H}} \tilde{\qV}_t \mathbf{E}_{i}\mathbf{\hat{h}}_k\mathbf{E}_{i}^{\rm{H}}  }{\Bigl(\sum_{j\in \mathcal{L}_k} |\mathbf{\hat{h}}_k^{\rm{H}} \tilde{\qV}_t \mathbf{E}_{j}|^2 + \sigma^2\Bigl)^2} \Biggl) 
\end{align}	
\end{small}

\vspace{-2mm}

\hrulefill

\vspace{-4mm}

\end{figure*}

\textbf{Search direction and update:} 
The search direction $\boldsymbol{\eta}_t$ at iteration $t$ is computed using a mini-batch of data, and given by $\boldsymbol{\eta}_t = - {\rm{grad}}_{\tilde{\qV}_t} \mathcal{L}_\rho(\tilde{\qV}, \boldsymbol{\lambda}, \boldsymbol{\kappa})$. The next iterate $\tilde{\qV}_{t+1}$ is computed by taking a step size $\alpha_t$ along $\boldsymbol{\eta}_t$, followed by a retraction to ensure the update remains on $\mathcal{M}$. The update rule is given by $ \tilde{\qV}_{t+1} = R_{\tilde{\qV}_t}(\alpha_t \boldsymbol{\eta}_t)$, where  $R_{\tilde{\qV}_t}(\cdot)$ is the retraction operation that maps the point from the tangent space back onto $\mathcal{M}$. The $\alpha_t$ is crucial for convergence and often follows a decaying strategy. It is defined as $\alpha_t = \frac{\alpha_0}{1 + \alpha_0 \lambda_t},$ where $\alpha_0$ is the initial step size and $\lambda$ is the decay parameter. This allows for larger steps at the start and gradually smaller steps as the algorithm converges. It ensures that the updates respect the manifold's geometry while efficiently guiding the iterates toward the optimal solution. The algorithm achieves faster iterations by applying stochastic updates, particularly when handling large datasets.

\textbf{Updating the Lagrange multipliers}:
Once $\tilde{\qV}$ is optimized, $\{\boldsymbol{\lambda}, \boldsymbol{\kappa}\}$ are updated to indicate progress in fulfilling the constraints. At iteration $t$, the updating rule for $\vartheta \in \{\lambda_n, \kappa_k\}$ is given as $\vartheta^{t+1} = \text{clip}_{[\vartheta^{\min}, \vartheta^{\max}]}\left(\vartheta^{t} + \rho_t \Psi(\tilde{\qV}_{t+1})\right)$, where $\Psi \in \{u_n, c_k\}$, $\rho_t > 0$ is a penalty parameter, and clipping confines $\vartheta^{t+1}$ to a predetermined range, i.e.,  $[\vartheta^{\min}, \vartheta^{\max}]$ \cite{liu2020simple}. This prevents the multipliers from increasing indefinitely, ensuring that the optimization process is consistent and regulated.

At each iteration $t$, SGALM algorithm generates a candidate solution $\mathcal{L}_{\rho}(\mathbf{\tilde{V}}_{t+1}, \boldsymbol{\lambda}_t, \boldsymbol{\kappa}_t) \leq \mathcal{L}_{\rho}(\mathbf{\tilde{V}}_t, \boldsymbol{\lambda}_t, \boldsymbol{\kappa}_t) + \epsilon_t$ with an infinite sequence $\{\epsilon_t\}$ that converges to zero, yielding a global minimizer for $\mathbf{P}$. This monotonically decreasing nature and the upper constraint imposed on the objective function ensures convergence \cite{zargari2024riemannian}. With $T$ iterations for convergence, the overall complexity is estimated as $\mathcal{O}(T (MK+M K^3))$ \cite{liu2020simple, zargari2024riemannian}.

\section{Simulation Results}\label{sim}
Simulation examples are presented to evaluate the performance of SGALM. The BS is placed at $\{0,0\}$. The users and targets are randomly distributed within circular regions centered at $\{40,10\}$ and $\{20,0\}$, respectively, with a radius of \qty{10}{\m}. The sensing directions from the BS to the targets are set at \qty{-65}{\degree}, \qty{-45}{\degree}, \qty{30}{\degree}, and \qty{60}{\degree}. The simulation consists of \num{e3} Monte Carlo trials. Table \ref{table-notations} presents the simulation parameters unless otherwise stated.

\begin{table}[t]\vspace{-1mm}
\renewcommand{\arraystretch}{1.0}
\centering
\caption{Simulation and algorithm parameters.}\vspace{-2mm}
\label{table-notations}
\begin{tabular}{c c c c}    
\hline
\textbf{Parameter}& \textbf{Value} & \textbf{Parameter}& \textbf{Value}\\  \hline \hline
$f_c$ & \qty{54}{\GHz}  & $\Omega_{n}^{\thr}$  & \qty{10}{\dB m}  \\ 
$M$ & \num{257}  &  $\sigma^2$  & \qty{-90}{\dB m} \\ 
$K$ & \num{2} & $p_{\rm{max}}$  & \qty{30}{\dB m} \\ 
$N$ & \num{4} & $\{\vartheta^{\min}, \vartheta^{\max}\}$ & \{\num{0}, \num{100}\} \\ 
$\mathcal{R}_k^{\thr}$ & \qty{15}{bps/\Hz} & $\rho_0$ & \num{1}\\ \hline
\end{tabular}
\vspace{-3mm}
\end{table}

\begin{figure}[!t]\vspace{-2mm}
    \centering 
    \includegraphics[width=65mm, height=55mm]{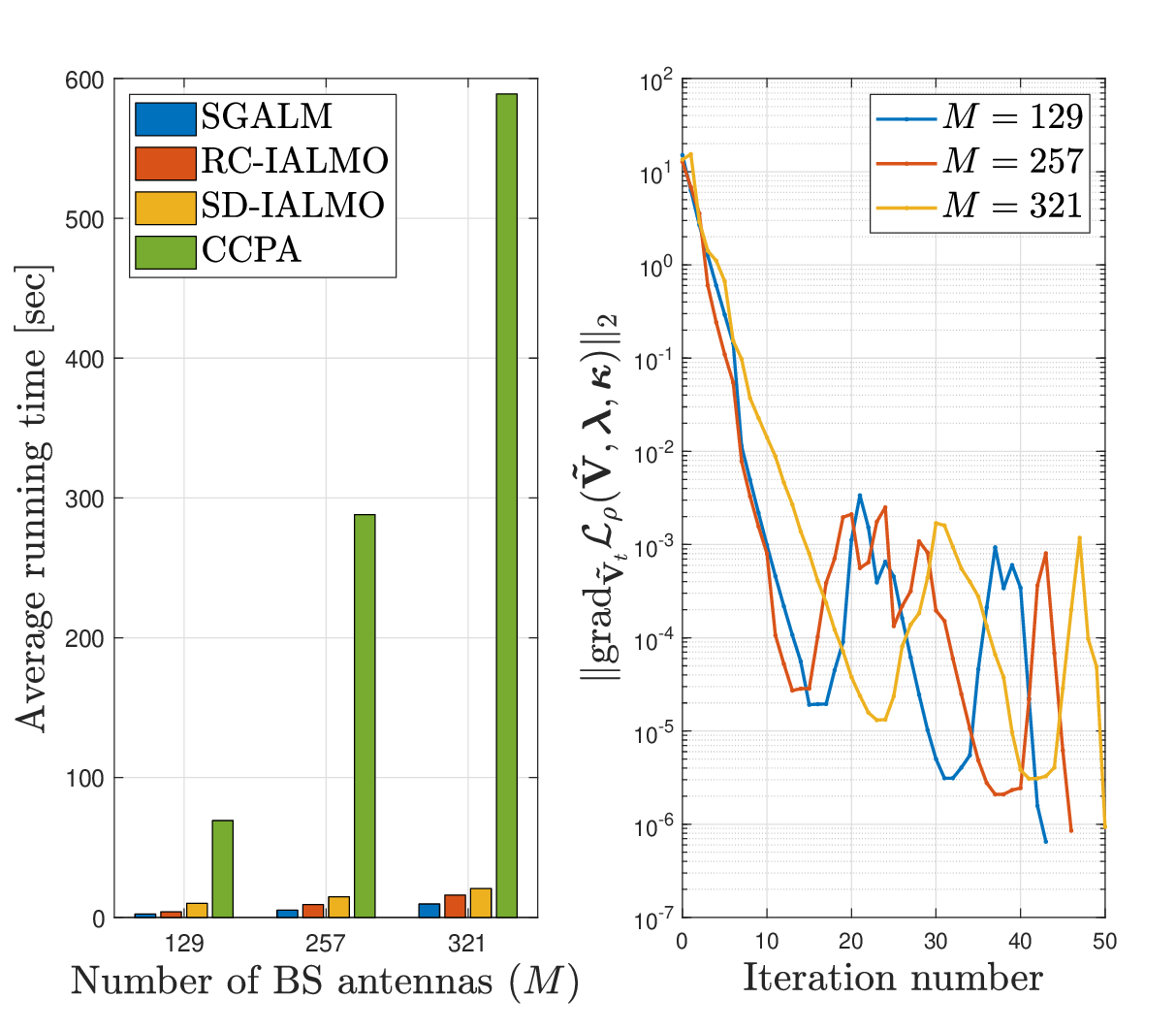}\vspace{-4mm}
    \caption{{The algorithm execution time (left) and convergence rate (right) versus the number of BS antennas.}}
    \label{fig_RunTime}\vspace{-3mm}
\end{figure}

Proposed SGALM is compared to an iterative convex-concave procedure algorithm (CCPA)-based benchmark \cite{Diluka2024NF}. This solves $\mathbf{P}$ using SDR and SCA iteratively \cite{Diluka2024NF}. In particular, $\qW_k = \qw_k \qw_k^{\rm{H}}$ and $\qS_n = \qs_n \qs_n^{\rm{H}}$ are defined, where $\qW_k$ and $\qS_n$ are semidefinite matrices with $\text{Rank}(\qW_k) = 1$. Then, $\mathbf{P}$ is reformulated as a conventional semi-definite problem (SDP) by relaxing the rank-one constraint \cite{Diluka2024NF}. The SDP problem is solved using the CVX tool. The relaxed rank one constraint is imposed using Gaussian randomization \cite{boyd2004convex}.

{Fig.~{\ref{fig_RunTime}} shows the average running times of SGALM and CCPA (left) and the convergence rate of the SGALM algorithm (right).} These are from Matlab simulations on an Intel\textsuperscript{\textregistered} Core\textsuperscript{\texttrademark} i7 processor at \qty{2.50}{\GHz}. In running time, we also compare SGALM with the steepest descent (SD)-IALMO and RC-IALMO, using different update strategies. SD-IALMO uses the basic steepest descent, while SGALM employs stochastic gradient updates for potentially faster convergence. Conversely, RC-IALMO utilizes information from previous iterations to update the search direction. As Fig.~\ref{fig_RunTime} indicates, the execution times for all algorithms increase with $M$. However, SGALM/IALMO consistently outperforms CCPA, significantly reducing average running time regardless of the number of BS antennas. For instance, with $M=\num{257}$, SGALM is $56$ times faster than CCPA.

The primary reasons are: (1) CCPA searches in the Euclidean space of dimensions $MKN$, while SGALM searches on a reduced manifold with $(M+1)(K+N)$ dimensions, lowering complexity; (2) SGALM directly reformulates the non-convex problem $(\mathbf{P})$  as an MO  problem without approximation; and (3) SGALM aggregates all beamforming vectors into a single variable, efficiently handling large-scale problems. In contrast, CCPA’s reliance on SCA-SDR approximations and matrix operations increases resource consumption as the antenna array grows, resulting in longer execution times. Moreover, SGALM outperforms SD-IALMO and RC-IALMO. This is due to the differences in their update methods.

{Fig.~{\ref{fig_RunTime}} (right) plots the norm of the gradient of the Lagrangian function, $\|{\rm{grad}}_{\mathbf{\tilde{V}}_t} \mathcal{L}_\rho(\mathbf{\tilde{V}}, \boldsymbol{\lambda}, \boldsymbol{\kappa})\|_2$. It evolves throughout iterations for varying numbers of antennas. Initially, the gradient norm declines rapidly for all settings, indicating that the optimization method approaches the optimal regions with lower gradient norms. As the iterations progress, the reduction becomes more slow, with more frequent fluctuations.  This shows the algorithm's step size and direction adjustment based on gradient guidance.}


\begin{figure}[!t]\vspace{-1mm}
    \centering 
    \includegraphics[width=65mm, height=55mm]{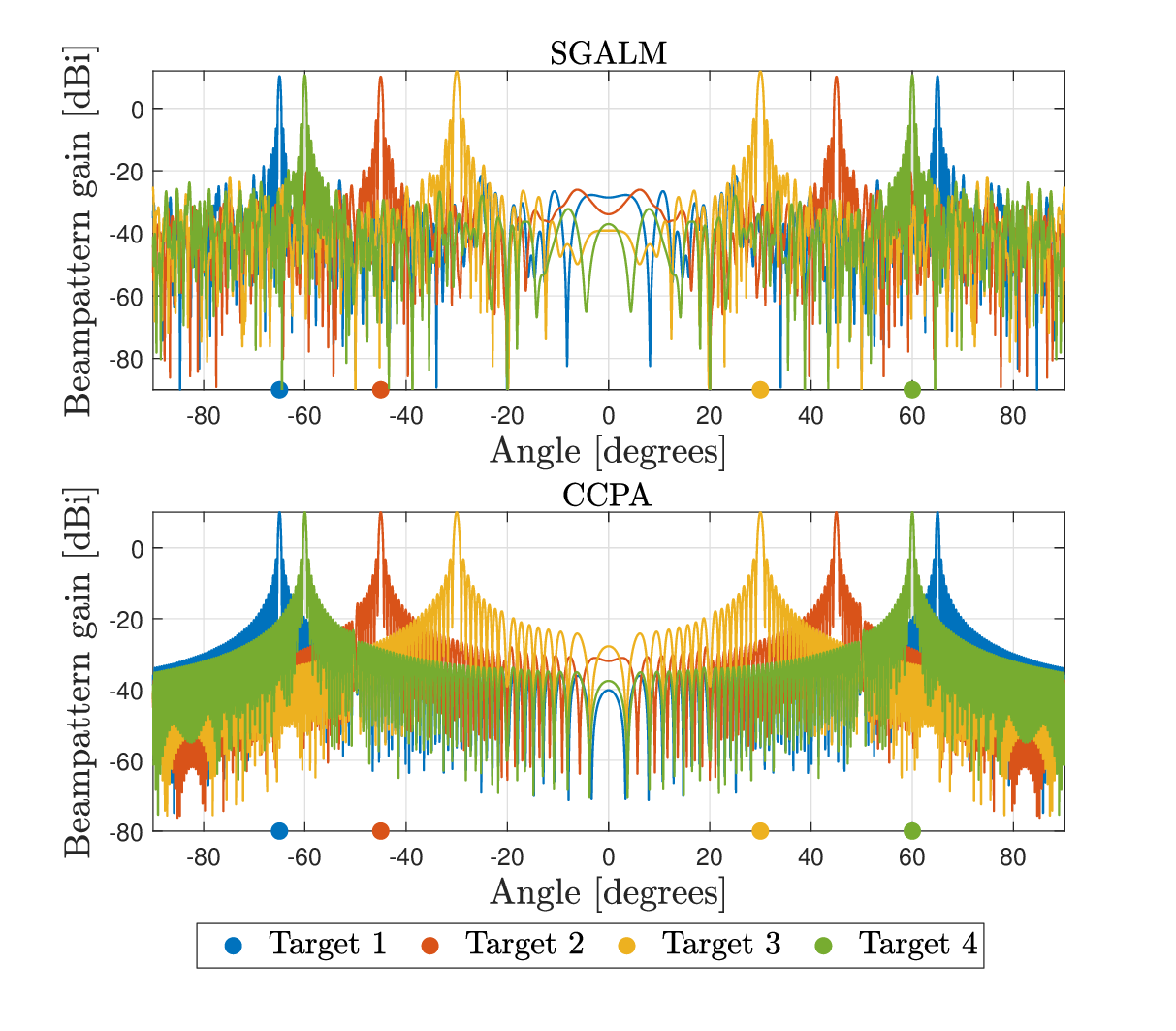}\vspace{-4mm}
    \caption{The beampattern gains for $M=\num{257}$ BS antennas.}
    \label{fig_BeamGain}\vspace{-3mm}
\end{figure}

\begin{figure}[!t]\vspace{-1mm}
    \centering 
    \includegraphics[width=65mm, height=55mm]{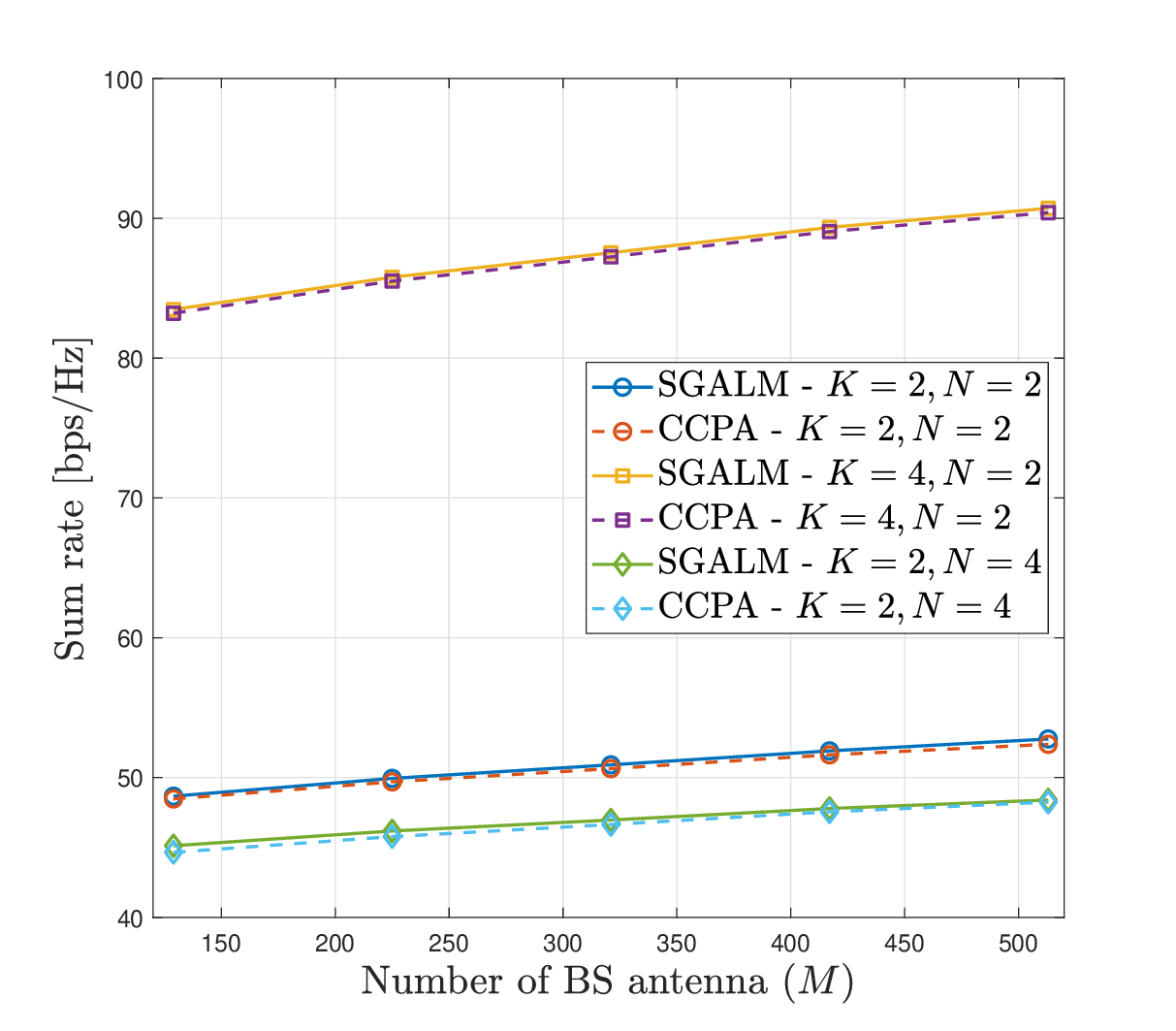}\vspace{-4mm}
    \caption{Sum rate versus the number of BS antennas.}
    \label{fig_SumRate}\vspace{-3mm}
\end{figure}

Fig.~\ref{fig_BeamGain} and Fig.~\ref{fig_SumRate} compare the transmit beampattern gains and sum rate performance of both algorithms.  They both exhibit similar beampattern gains and sum rate performance. It emphasizes their efficiency in focusing radiated power in targeted directions, which is necessary for maximizing spatial filtering and mitigating interference. Nevertheless, SGALM has lower complexity and is faster than CCPA, a critical requirement for ELAA-ISAC systems.

\begin{figure}[!t]\vspace{-1mm}
    \centering 
    \includegraphics[width=65mm, height=55mm]
{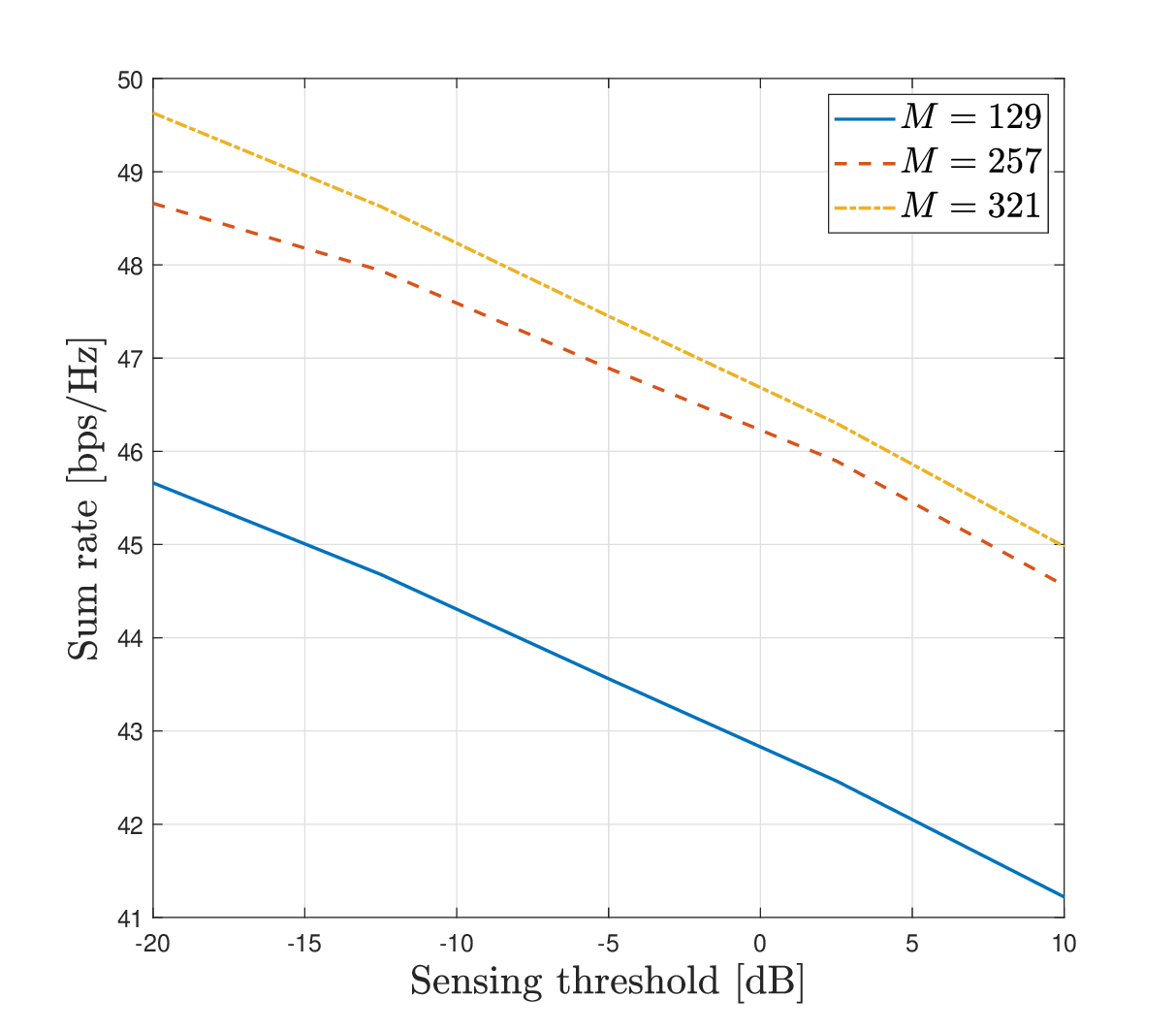}\vspace{-4mm}
    \caption{Sum rate versus the sensing beampattern gain threshold.}
    \label{fig_SumRateVsSensingThr}\vspace{-3mm}
\end{figure}

Fig.~\ref{fig_SumRateVsSensingThr} investigates the trade-off between C\&S tasks.  It thus plots the sum rate as a function of the sensing beampattern gain threshold, $\Omega_{n}^{\thr}$, for various values of $M = \{129, 257, 321\}$. The figure highlights an inverse relationship between C\&S: as the demand for sensing beampattern gain increases, the sum rate decreases. This trade-off arises from the concurrent operation of both functions. As sensing performance improves, interference on the user side intensifies, and more resources are allocated to sensing, reducing communication performance.

\section{Conclusion}
This study developed a low-complexity beamforming algorithm, SGALM,  for simultaneous multi-target detection and communication in  ELLA ISAC systems. SGALM exploits the beam-focusing effect in NF systems when the BS  generates multiple beams, enabling it to maximize communication sum rates and achieve sensing gain. Although the complexity of this problem grows exponentially with the number of antennas, SGALM reduces the search space from \(MKN\) dimensions to a manifold of \((M+1)(K+N)\), significantly cutting complexity. Consequently, it is notably faster than the conventional CCPA,  SD-IALMO, and RC-IALMO algorithms. SGALM  uses more efficient search directions and achieves faster convergence through the Riemannian SGD method. This works well for ELLA-ISAC systems ($M\geq 100).$

Future work can explore the feasibility of using dedicated sub-antenna arrays for C\&S and applying SGALM across various ISAC topologies. Channel estimation and beam training also present promising research directions.

\bibliographystyle{IEEEtran}
\bibliography{IEEEabrv,ref}
\end{document}